\documentclass[aps,prl,showpacs,twocolumn,amsmath,amssymb,
tightenlines,epsfig]{revtex4}

\usepackage{graphicx,color}
\usepackage{subfigure}

\begin{document}

\preprint{\begin{tabular}{l}
\hbox to\hsize{December, 2002 \hfill KAIST-TH 2003/04}\\
\hbox to\hsize{hep-ph/0212092 \hfill MCTP-03-17}\\
\hbox to\hsize{\hfill MADPH-03-1327}\\
\end{tabular} }

\title{$B_d \rightarrow \phi K_S$ CP asymmetries as an important 
probe of supersymmetry } 

\author{G.L. Kane}
\affiliation{Randall Laboratory, University of Michigan \\
Ann Arbor, MI 48109, U.S.A.}

\author{P. Ko}
\altaffiliation{On leave of absence from 
Department of Physics, KAIST, Daejon 305-701, Korea}
\affiliation{Randall Laboratory, University of Michigan \\
Ann Arbor, MI 48109, U.S.A.}

\author{C. Kolda}
\affiliation{Department of Physics, University of Notre Dame \\ 
Notre Dame, IN 46556, U.S.A.}

\author{Jae-hyeon Park}
\affiliation{Department of Physics, KAIST \\ Daejon 305-701, Korea}

\author{Haibin Wang}
\affiliation{Randall Laboratory, University of Michigan \\
Ann Arbor, MI 48109, U.S.A.}

\author{Lian-Tao Wang}
\affiliation{ Department of Physics, Univ. of Wisconsin  \\
1150 University Ave.
Madison, WI 53706, U.S.A.}

\date{\today}

\begin{abstract}
The decay $B_d \rightarrow \phi K_S$ is a special probe of 
physics beyond the Standard Model (SM), since it has no SM tree level 
contribution. 
Motivated by recent data suggesting a deviation from the SM for its  
time-dependent CP asymmetry, we examine supersymmetric explanations.  
Chirality preserving contributions are generically small, 
unless gluino 
is relatively light.
Higgs contributions are also too small to explain a large asymmetry.
Chirality flipping $LR$ and $RL$ gluino contributions 
actually can provide sizable effects 
without conflict with all related results. 
We discuss how various insertions can be distinguished, 
and argue 
the needed sizes  of mass insertions are reasonable.

\end{abstract}

\pacs{12.60.Jv, 11.30.Er}

\maketitle


$B\rightarrow \phi K$ is a powerful testing ground for new physics. 
Because it is loop suppressed in the standard model (SM), 
this decay is very sensitive to
possible new physics contributions to $b\rightarrow s s \bar{s}$, 
a feature not shared by other charmless $B$ decays. 
Within the SM, it is dominated by the  QCD penguin diagrams 
with a top quark in the loop. 
Therefore the time dependent CP asymmetries are essentially 
the same as those in $B\rightarrow J/\psi K_S$: $\sin 2 \beta_{\phi K_S} 
\simeq \sin 2 \beta_{J/\psi K_S} + O(\lambda^2)$ \cite{worah}. 

Recently both BaBar and Belle reported the branching ratio and CP asymmetries
in  the  $B_d \rightarrow \phi K_S$ decay: 
\begin{eqnarray}
{\cal A}_{\phi K} (t) & \equiv &
{{\Gamma (\overline{B}^0_{\rm phys} (t) \rightarrow \phi K_S ) - 
 \Gamma (      B^0_{\rm phys} (t) \rightarrow \phi K_S )   } \over
{\Gamma (\overline{B}^0_{\rm phys} (t) \rightarrow \phi K_S ) + 
 \Gamma (      B^0_{\rm phys} (t) \rightarrow \phi K_S )   }}
\nonumber  \\
& = & - C_{\phi K} \cos ( \Delta m_B t )
  + S_{\phi K} \sin ( \Delta m_B t ),
\end{eqnarray}
where $C_{\phi K}$ and $S_{\phi K}$ are given by
\begin{equation}
C_{\phi K}  = 
{ 1 - | \lambda_{\phi K} |^2 \over 1 + | \lambda_{\phi K} |^2 } , ~~~~~
{\rm and}~~~~~
S_{\phi K}  =  
{ 2~ {\rm Im} \lambda_{\phi K} \over 1 + | \lambda_{\phi K} |^2 } ,
\end{equation}
with 
\begin{equation}
\lambda_{\phi K} \equiv - e^{ - 2 i (\beta + \theta_d )} 
{\overline{A} ( \overline{B}^0 \rightarrow \phi K_S ) \over 
A ( B^0 \rightarrow \phi K_S ) } ,
\end{equation}
and the angle $\theta_d$ represents any new physics contributions to the 
$B_d - \overline{B_d}$ mixing angle. They find 
$\sin 2 \beta_{\phi K} = S_{\phi K} = -0.39 \pm 0.41$ \cite{babar,belle}, 
which is a 2.7 $\sigma$ deviation 
from the SM prediction: $\sin 2 \beta_{J/\psi K_S} = 0.734 \pm 0.054$ 
\cite{nir}. On the other hand, the measured branching ratio 
$\sim (8-9) \times 10^{-6}$ is not far from the SM prediction. 
The direct CP asymmetry in $B_d \rightarrow \phi K_S$ is also reported by the 
Belle collaboration: 
$C_{\phi K_S} = 0.56 \pm 0.43$.  

In this paper, we wish to explain the deviation of $\sin 2\beta_{\phi K_S}$ 
from  $\sin 2 \beta_{J/\psi K_S}$
within general SUSY models with $R-$parity conservation. (See Refs.~
\cite{moroi} for related works.) 
There are basically two interesting classes of modifications:
(i) gluino-mediated $b\rightarrow s q \bar{q}$  with
$q = u,d,s,c,b$ induced by flavor mixings in the down-squark sector, and 
(ii) Higgs mediated $b\rightarrow s s \bar{s}$ in the large $\tan\beta $ 
limit ($\propto \tan^3 \beta$ at the amplitude level). 
In the following, we analyze the effects of these two mechanisms on 
the branching ratio of $B_d \rightarrow \phi K_S$, $S_{\phi K}$, 
$C_{\phi K}$, $B\rightarrow X_s \gamma$ and its direct CP asymmetry, 
$B_s^0 - \overline{B_s^0}$ mixing, and  
the correlation of the Higgs mediated $b\rightarrow s s \bar{s}$ 
transition with $B_s \rightarrow \mu^+ \mu^-$. 
We will  deduce that $LL$ and $RR$ insertions give effects 
too small to cause an observable deviation between $S_{\phi K}$ and 
$S_{J/\psi K}$, unless the gluino and squarks have masses close to the 
current lower bounds.
Once the existing CDF limit on $B_s \rightarrow \mu^+ \mu^-$ is imposed on 
the Higgs mediated $b\rightarrow s s \bar{s}$, that also provides too small 
an effect. On the other hand, the down-sector $LR$ and $RL$ insertions can 
provide a sizable deviation in a robust way consistent with all other data.


In the general MSSM, the squark and the quark mass matrices are not 
diagonalized simultaneously. There will be gluino mediated FCNC of strong 
interaction strength, which may even exceed existing limits.
It is customary to rotate the effects
so they occur in squark propagators rather than in couplings, and to 
parametrize them in terms of dimensionless parameters.
We work in the usual mass insertion approximation (MIA) \cite{mia}, 
where the flavor mixing $j \rightarrow i$ in the down type squarks 
associated with $\tilde{q}_B$ and $\tilde{q}_A$ is parametrized by 
$( \delta_{AB}^d )_{ij}$. More explicitly, 
\[ 
( \delta_{LL}^d )_{ij} = \left( V_L^{d \dagger} M_{Q}^2 V_L^d \right)_{ij}  
/ \tilde{m}^2, ~~~{\rm etc.} 
\] 
in the super CKM basis where the quark mass matrices are diagonalized by
$V_L^d$ and $V_R^d$, and the squark mass matrices are rotated in the same
way. Here $ M_{Q}^2$, $M_{D}^2$ and $M_{LR}^2$ are squark mass matrices, 
and $\tilde{m}$ is the average squark mass. Because we are considering
$b\to ss\bar s$ transitions, we have 
$(i,j)=(2,3)$.

Since the decay $B_d \rightarrow \phi K_S$ is dominated by SM QCD penguin 
operators, the gluino-mediated QCD penguins may be significant 
if the gluinos and squarks are not too heavy. 
For $LL$ and $RR$ mass insertions, the gluino-mediated diagrams can
contribute to a number of operators already present in the SM, such as
those with LH flavor-changing 
currents ($O_{3,\ldots,6}$) and the magnetic and
chromomagnetic operators ($O_{7\gamma}$, $O_{8g}$). On the other hand
$LR$ and $RL$ insertions do not generate $O_{3,\ldots,6}$. (All operators
are explicitly defined in \cite{bbns}.)
But $RR$ and $RL$ insertions also generate operators with
RH flavor-changing currents and so we must consider the operators
$\widetilde{O}_{3,\ldots,6}$, $\widetilde{O}_{7\gamma}$ and
$\widetilde{O}_{8g}$ which are derived from the untilded operators by
the exchange $L\leftrightarrow R$. 
In the SM, the $\widetilde{O}_{3,\ldots,6}$ operators are absent
and the $\widetilde{O}_{7\gamma,8g}$ operators are suppressed by
light quark masses, $m_s$. But in SUSY, $\widetilde{O}_{7\gamma,8g}$
in particular can be enhanced by $\tilde{m}/m_s$ with respect to
the SM, making them important phenomenologically.  Of course,
these gluino induced FCNC could also affect  $B\rightarrow J/\psi K_S$
in principle. But this gold-plated mode for $\sin 2\beta$ is dominated
by a tree level diagram in the SM, and the SUSY loop corrections are 
negligible. 

We calculate the Wilson coefficients corresponding to each of these 
operators at the scale $\mu\sim \tilde{m}\sim m_W$; explicit expressions are
in Ref.~\cite{kkkpww}.   We evolve the Wilson coefficients to 
$\mu \sim m_b$ using the appropriate renormalization group (RG)
equations \cite{rg}, 
and calculate the amplitude for $B\rightarrow \phi K$ using the recent 
BBNS approach~\cite{bbns}  for estimating the hadronic matrix elements.
  
The explicit expressions for $\Gamma(B\to\phi K_S)$
can be found in Refs.~\cite{kkkpww,bbns}.
In the numerical analysis presented here, we 
fix the SUSY parameters to be $m_{\tilde{g}} = \tilde{m} = 400$ GeV.
In each of the mass insertion scenarios to be discussed, we vary
the mass insertions over the range $\left|\delta^d_{AB}\right|\leq
1$ to fully map the parameter space.
We then impose two important experimental constraints.
First, we demand that the predicted branching ratio for inclusive
$B\to X_s\gamma$ fall within the range
$2.0 \times 10^{-4} < B ( B\rightarrow X_s \gamma ) < 4.5 \times 10^{-4}$,
which is rather generous 
in order to  allow for various theoretical uncertainties. 
Second, we impose the current lower limit on
$\Delta M_s > 14.9$ ps$^{-1}$. 

A new CP-violating phase in $( \delta^d_{AB} )_{23}$ will also 
generate CP violation in $B\rightarrow X_s \gamma$. The current 
data on the direct CP asymmetry 
$A_{\rm CP}^{b\rightarrow s \gamma}$ from CLEO 
is \cite{cleo} 
$A_{\rm CP}^{b\rightarrow s \gamma} = (-0.079 \pm 0.108 \pm 0.022)
(1.0 \pm 0.030)$, which is  not particularly constraining.  
Within the SM, the predicted CP asymmetry is less than $\sim 0.5 \%$, so 
a larger asymmetry would be a clear indication of new physics \cite{kn}.   
Remarkably,  despite the well-known    
importance of $B\rightarrow X_s \gamma$ for constraining
$( \delta_{AB}^d )_{23}$, we will  find in the following that
the $B\rightarrow \phi K$ branching ratio provides an independent 
constraint on the $LR$ and $RL$ insertions which is already 
as strong as that derived from $B\rightarrow X_s \gamma$. Where
relevant, we will show our predictions for 
$A_{\rm CP}^{b\rightarrow s \gamma}$.

We begin by considering the case of a single $LL$ 
mass insertion: $(\delta^d_{LL})_{23}$. Our results will hold equally
well for a single $RR$ insertion. 
Though the $LL$ insertion generates all of the operators
$O_{3,\ldots,6,7\gamma,8g}$,
qualitatively there are substantial cancellations between 
their SUSY contributions and so
the effect of the $LL$ insertion is significantly diluted. 
Scanning over the parameter space as discussed previously,
we find that $S_{\phi K} > 0.5$
for $m_{\tilde{g}} = \tilde{m} = 400\,$GeV  and 
for any value of $| ( \delta_{LL}^d )_{23} | \leq 1$,
the lowest values being achieved only for very large $\Delta M_s$. 
If we lower the gluino mass down to $250\,$GeV,  $S_{\phi K}$ can shift down 
to $\sim 0.05$, but  only in a small corner of parameter space.
Similar results hold for a single $RR$ insertion.   Thus we 
conclude that the effects of the $LL$ and $RR$ insertions on  
$B\rightarrow X_s \gamma$ and $B\rightarrow \phi K$ 
are not sufficient to generate a  (large) negative $S_{\phi K}$,
unless  gluino and squarks are relatively light.  
Nonetheless, their effects 
on $B_s - \overline{B_s}$ mixing could still be large  and
observable, providing a clear signature for $LL$ and $RR$ mass
insertions (see Ref.~\cite{kkkpww} for details). 

Next we consider the case of a single $LR$ insertion.
Scanning over the parameter space and imposing the constraints 
from $B\to X_s\gamma$ and $\Delta M_s$, we find 
$| ( \delta_{LR}^d )_{23} | \lesssim 10^{-2}$.  This is, however,
large enough to significantly affect $B\to\phi K_s$,
both its branching ratio and CP asymmetries, through the
contribution to $C_{8g}$.
\begin{figure}
{\includegraphics[width=4cm]%
{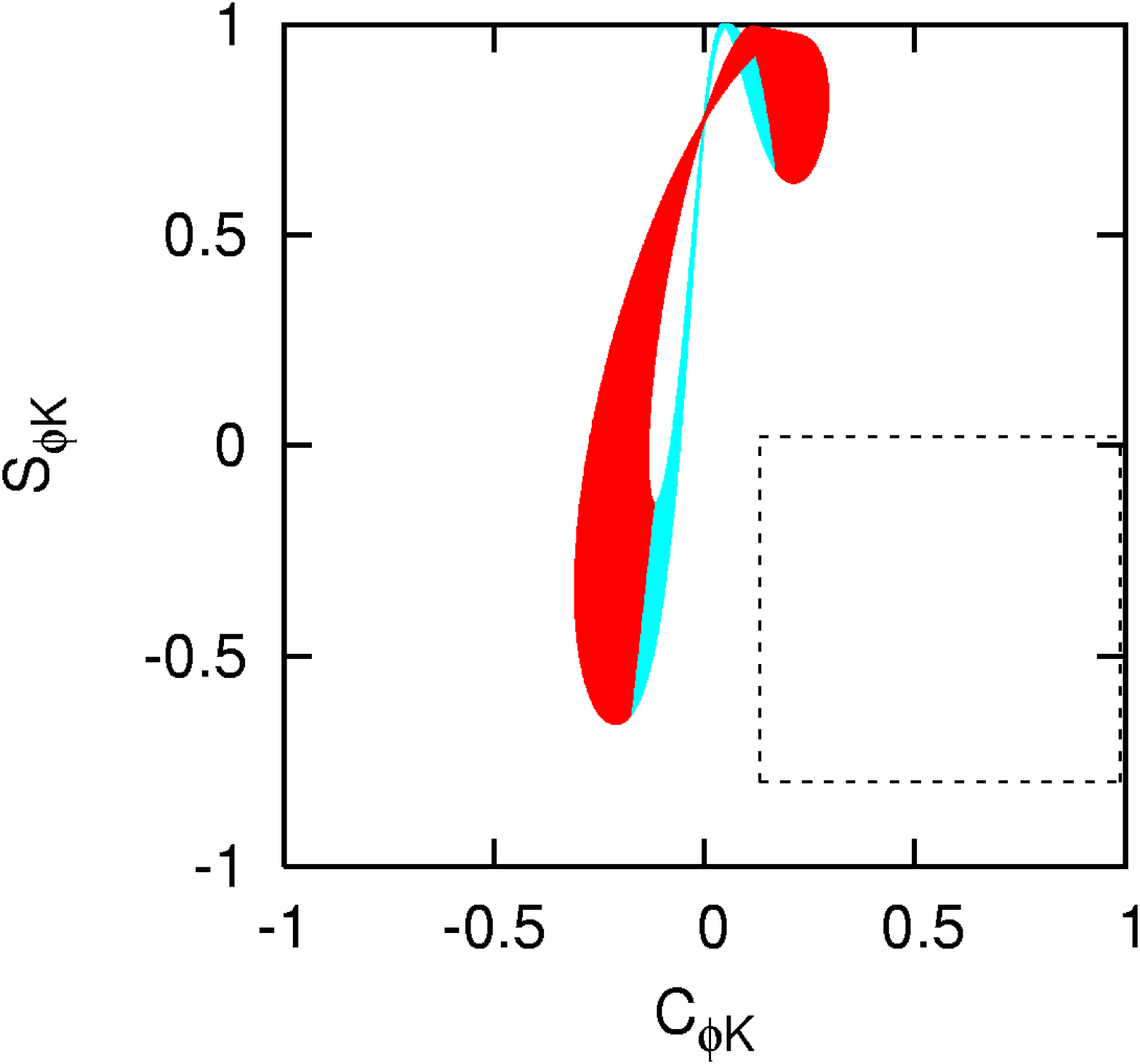}}
{\includegraphics[width=4cm]
{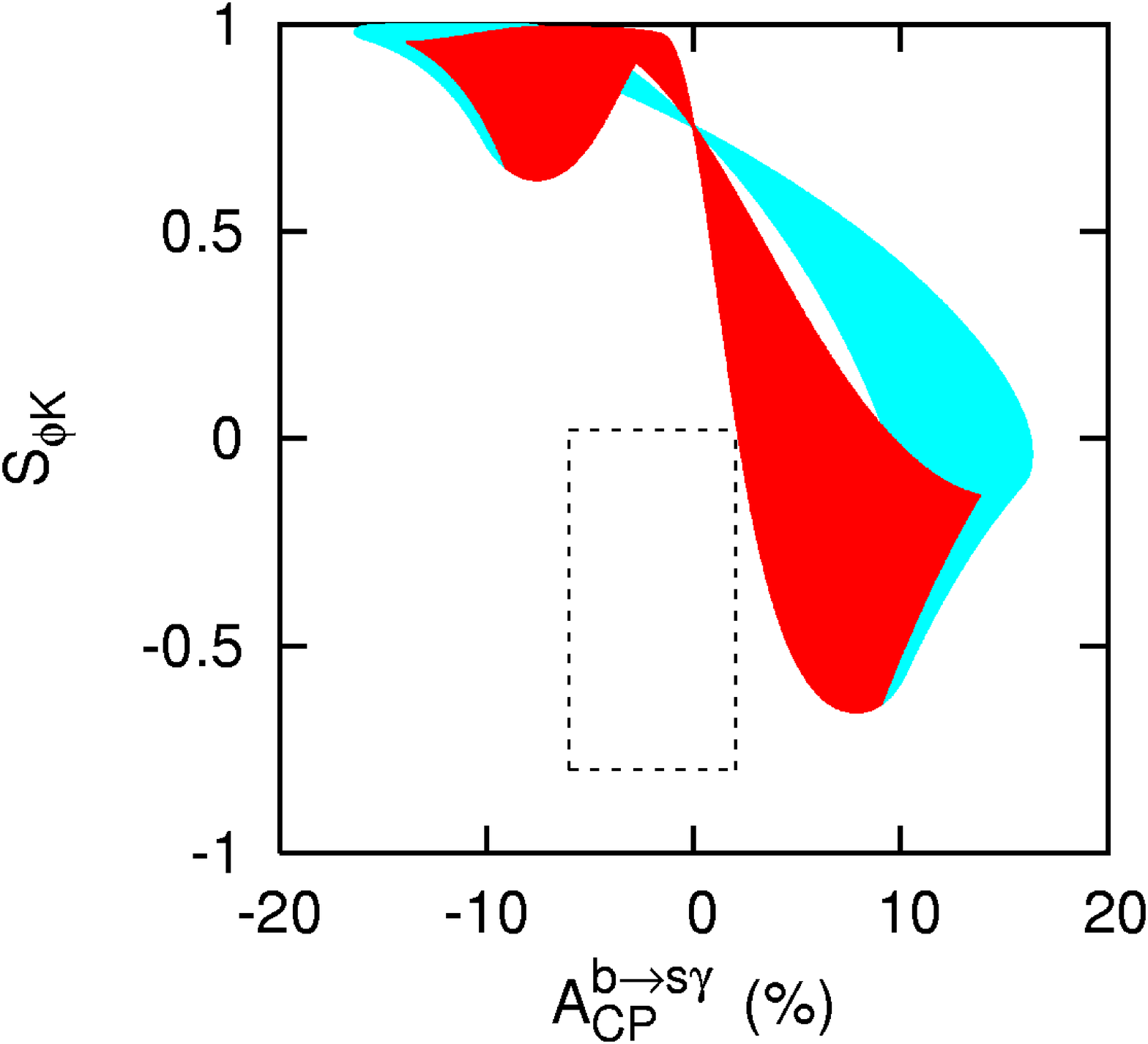}}
 \caption{
The allowed region in the plane of  
(a) $S_{\phi K}$ and $C_{\phi K}$, and  
(b) $S_{\phi K}$ and $A_{\rm CP}^{b\rightarrow s \gamma}$,
for the case of a single $LR$ insertion,  with
$m_{\tilde{g}} =\tilde m= 400\,$GeV.
The dotted boxes show the current $1\sigma$ expermental bounds, and 
the light (cyan) regions correspoind to 
$B ( B_d \rightarrow \phi K^0 ) > 1.6 \times 10^{-5}$.
 }
\end{figure}
In Fig.~1(a), we show the correlation between $S_{\phi K}$ and $C_{\phi K}$.
Since the  $LR$ insertion can have a large effect on the
CP-averaged branching ratio for $ B\rightarrow \phi K$ 
we further impose that $B ( B\rightarrow \phi K) < 1.6 \times 10^{-5}$ 
(which is twice the experimental value)
in order to include theoretical 
uncertainties  in the BBNS approach 
related to hadronic physics. Surprisingly, we see that the 
$B\rightarrow \phi K$ branching ratio constrains 
$ ( \delta_{LR}^d )_{23} $  just as much as $B\rightarrow X_s \gamma$.

Studying Fig.~1, we can learn much about the viability and
testability of scenarios with a single $LR$ insertion. For one thing,
we can get negative $S_{\phi K}$, but only if $C_{\phi K}$ is also 
negative (this is excluded by the present data, but only at the
$1.4\sigma$ level).
Conversely, a positive $C_{\phi K} $ implies that $S_{\phi K} > 0.6$. 
This correlation between $S_{\phi K}$ and $C_{\phi K}$ can be 
tested at B factories in the near future  given better statistics.  
The correlation between $S_{\phi K}$ and the direct CP asymmetry in 
$B \rightarrow X_s \gamma$ ($\equiv A_{\rm CP}^{b\rightarrow s \gamma}$) 
is shown in Fig.~1(b). We find $A_{\rm CP}^{b\rightarrow s \gamma}$ 
becomes positive for a negative $S_{\phi K}$, while a negative 
$A_{\rm CP}^{b\rightarrow s \gamma}$ implies that $S_{\phi K} > 0.6$.  
It is also clear from the figure that  
the present CLEO bound on $A_{\rm CP}^{b\rightarrow s \gamma}$ 
does not significantly constrain the $LR$ model. 
Finally, we also find that  
the deviation of $B_s - \overline{B}_s$ mixing from the SM prediction 
is small after imposing the $B_d \rightarrow X_s \gamma$ and 
$B_d \rightarrow \phi K_S $ branching ratio constraints. 
Thus we conclude that a single $LR$ insertion can describe the
data on $S_{\phi K}$ but only if the measurement of $C_{\phi K}$
shifts to negative values. This scenario can then be tested by
measuring a positive direct CP asymmetry in $B\to X_s\gamma$ and
$B_d$-$\bar B_d$ mixing consistent with the SM. 


The last single mass insertion case is that of $RL$.  
We find that the $RL$ operator most conveniently describes the current
data. For example,  
in Figs.~2(a) and (b), we show the correlation of $S_{\phi K}$ with 
$B ( B\rightarrow \phi K )$ and $C_{\phi K}$, respectively, in the 
$RL$ dominance scenario.  We find that  $S_{\phi K}$ can take almost 
any value between $-1$ and $+1$ without conflict with the observed 
branching ratio for $B\rightarrow \phi K$.  We also find that
$C_{\phi K} $ can be positive for a negative $S_{\phi K}$, unlike 
in the $LR$ case. 
In particular, if we impose the $B(B_d\to \phi K^0)$ constraint, then
$|C_{\phi K}| >0.2$ for $S_{\phi K} < 0$, except in very small regions
of parameter space.  
This could be a useful check on the presence of the $RL$ operator.
But the $RL$ operator leaves SM predictions intact in two
important  observables. First, within the range of allowed
$(\delta^d_{RL})_{23}$, the $B_s$-$\bar B_s$ mixing receives only very
small contributions from SUSY. Second, because the $RL$ insertion
generates the $\widetilde{O}_{3,\ldots,6,7\gamma,8g}$ 
operators while the SM generates only the untilded
operators, there is no interference between the phases of the SUSY and
SM contributions, and so  $A_{\rm CP}^{b\to s\gamma} = 0$.   

The process $B\to X_s\gamma$ actually has a rather unconventional but
well-motivated structure in the $RL$ scenario. In 
Ref.~\cite{kane},  it was found that one can have a strong
cancellation between the SM and SUSY contributions to $C_{7\gamma}$ and
$C_{8g}$,  as is expected in the supersymmetric limit. Then the
observed branching ratio is due to the $\widetilde{C}_{7\gamma}$
operator, which  implied an  $RL$ insertion of about the same size  
as that needed here. Thus this earlier analysis motivates the $RL$ insertion
for the present case.

We have considered separate insertions so the effects of each can be seen
clearly.  
In many realistic SUSY models, combinations of various insertions 
will  play a role phenomenologically,  and of course we do not mean to 
imply that only single insertions should occur in nature.  
A particularly interesting case is one in which the $RL$ and $RR$ 
operators co-exist. While we do not study this model in detail here, 
we note that (as discussed in Ref.~\cite{kane}) 
if both $( \delta_{RR}^d )_{23}$ and $(\delta^d_{RL})_{23}$ are  nonzero, 
then $A_{\rm CP}^{b\rightarrow s \gamma}$ can arise as an interference 
between $\widetilde{C}_{7\gamma}$ and $\widetilde{C}_{8g}$. 
Then the $RR$ insertion could shift $\Delta M_s$ appreciably, while 
the $RL$ insertion gives the above result.

\begin{figure}
{\includegraphics[width=4cm]
{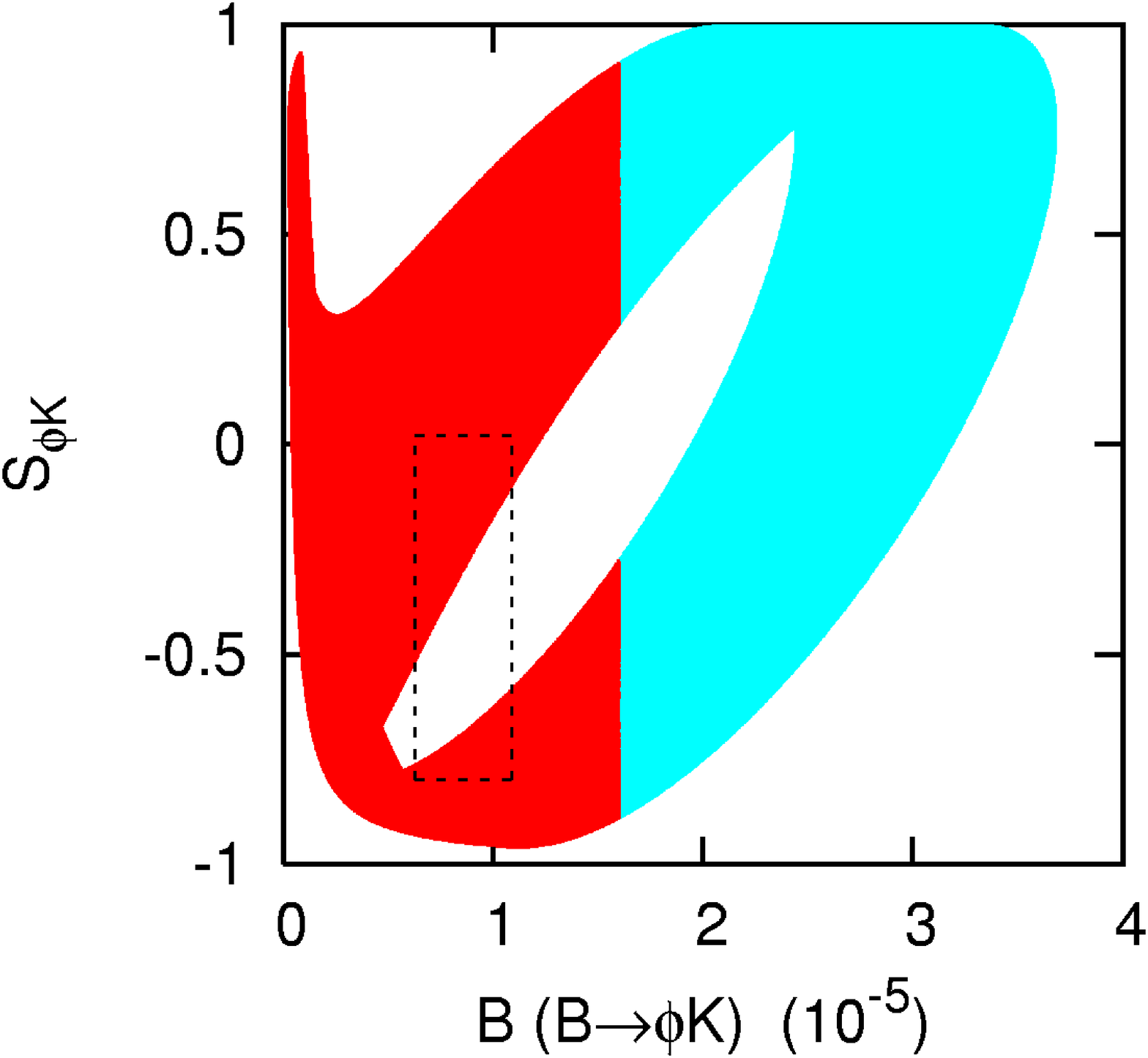}}
{\includegraphics[width=4cm]%
{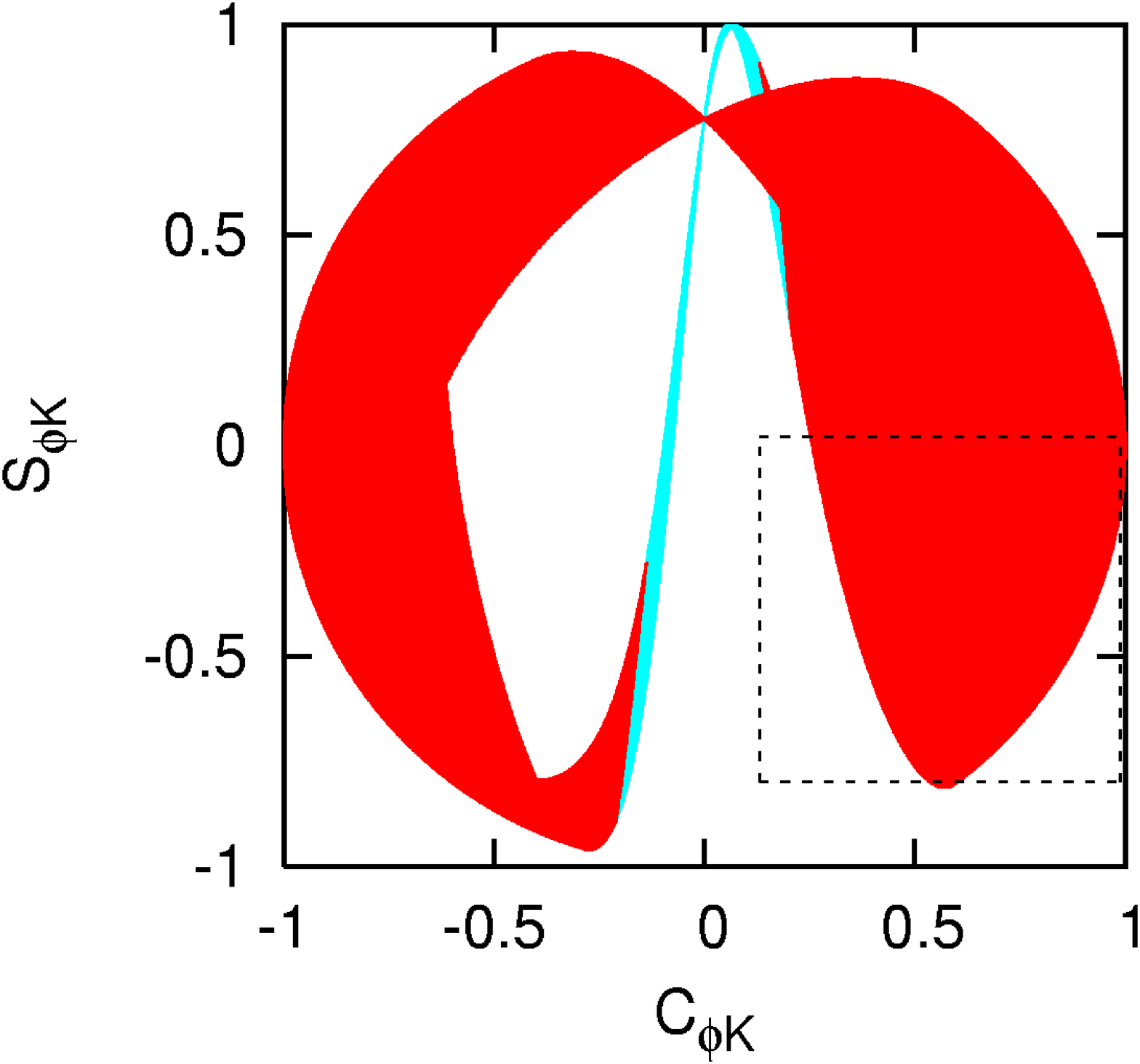}}
 \caption{
For the case of a single $RL$ insertion, we show the correlation
between (a) $S_{\phi K}$ and $B ( B\rightarrow \phi K )$, 
and (b) between $S_{\phi K}$ and $C_{\phi K}$. 
Current  $1\sigma$ limits are shown 
by the dotted boxes,  while the light (cyan) regions correspond to 
$B ( B_d \rightarrow \phi K^0 ) > 1.6 \times 10^{-5}$.
}
 \end{figure}


Finally we consider a completely different class of
contributions to   $b\to ss\bar s$.  
At large $\tan\beta$, FCNC's can also be mediated by exchanges of neutral 
Higgs bosons, as found in  $B\rightarrow X_s \mu^+\mu^-$ 
\cite{huang}, $B_{s,d}\rightarrow \mu^+ \mu^-$ \cite{kolda}, 
$\tau \rightarrow 3 \mu$, etc. \cite{Babu:2002et}.  Likewise, 
$b\rightarrow s s \bar{s}$ could be enhanced by neutral Higgs  exchange, 
which is flavor dependent since the Higgs coupling is proportional to 
the Yukawa couplings.
The effective coupling for $b\rightarrow s s \bar{s}$ is basically the same
with that for $B_s \rightarrow \mu^+ \mu^-$ up to a small difference 
between the muon and the strange quark masses (or their Yukawa couplings). 
Imposing the upper limit  on this decay from CDF \cite{cdf} during the 
Tevatron's Run I, 
$B( B_s \rightarrow\mu^+ \mu^- ) < 2.6 \times 10^{-6}$, 
we can derive a model-independent upper limit on this effective coupling.
We find that $S_{\phi K_S} $ cannot be smaller than 0.71 for such 
couplings,  so that we cannot explain the large deviation in
$S_{\phi K_S}$ with Higgs-mediated $b\rightarrow s s \bar{s}$  alone.

Now let us provide possible motivation for values of
$( \delta_{LR,RL}^d )_{23} \lesssim 10^{-2}$ that we find phenomenologically 
are needed  in order to generate $S_{\phi K}$.   In particular, 
at large $\tan\beta$ it is possible to have double mass insertions
which give sizable contributions to $( \delta_{LR,RL}^d )$. First a
$( \delta_{LL}^d )_{23}$ or $( \delta_{RR}^d )_{23}\sim 10^{-2}$ is
generated. The former can be obtained from renormalization group
running even if its initial value is negligible at the high scale.
The latter may be implicit in SUSY GUT models with large mixing in the
neutrino sector \cite{moroi}.
Alternatively, in models in which the SUSY flavor problem is resolved 
by an alignment mechanism using 
spontaneously broken flavor symmetries, or  by decoupling,  
the resulting $LL$ or $RR$ mixings in the $23$ sector could easily be of 
order $\lambda^2$ \cite{align,decoupling}.  However as discussed above, 
this size of the $LL$ and/or $RR$ insertions cannot explain
the measured CP asymmetry in $B_d \rightarrow \phi K_S$.
But  at large $\tan\beta$, the $LL$ and $RR$ insertions can
induce the $RL$ and $LR$ insertions needed for $S_{\phi K}$ through a  
double mass insertion \cite{bjkp}:
\[
( \delta_{LR,RL}^d )_{23}^{\rm ind} =  ( \delta_{LL,RR}^d )_{23} \times 
{ m_b ( A_b - \mu \tan\beta ) \over \tilde{m}^2 }.
\]
One can achieve $( \delta_{LR,RL}^d )_{23}^{\rm ind} \sim 10^{-2}$
if $\mu \tan\beta \sim 10^4\,$GeV, which could be natural if $\tan\beta$ is 
large (for which $A_b$ becomes irrelevant). 
Note that in this scenario both the $LL (RR)$ and $LR (RL)$ insertions 
would have the same CP violating phase, since the phase of $\mu$ here is
constrained by electron and down-quark electric dipole moments. 
Lastly, one can also construct string-motivated $D$-brane 
scenarios in which $LR$ or $RL$ insertions are $\sim 10^{-2}$ \cite{kkkpww}. 


In this letter, we considered several classes of  potentially important 
SUSY contributions to $B\rightarrow \phi K_S$  in order to 
see if a significant deviation in its time-dependent CP asymmetry 
$S_{\phi K_S}$ could arise from SUSY effects. 
The Higgs-mediated FCNC effects
and models based on the gluino-mediated $LL$ and $RR$ insertions 
both give contributions too small to alter $S_{\phi K}$ significantly. 
On the other hand, the gluino-mediated contribution with $LR$ and/or $RL$ 
insertions can lead to sizable negative in $S_{\phi K_S}$ (as reported 
experimentally) as long as 
$| ( \delta_{LR,RL}^d )_{23} | \sim 10^{-3} - 10^{-2}$.
As a byproduct, we found that nonleptonic $B$ decays such as $B\rightarrow
\phi K$ begin to constrain $| ( \delta_{LR,RL}^d )_{23} | $ 
as strongly as $B\rightarrow X_s \gamma$.  
Besides producing no measurable deviation in $B^0$-$\bar B^0$ mixing, 
the $RL$ and $LR$  operators generate definite
correlations among $S_{\phi K}$, $C_{\phi K}$ and 
$A_{\rm CP}^{b\rightarrow s \gamma}$, and our explanation for the negative 
$S_{\phi K}$ can be easily tested by measuring these other  observables. 
In particular $C_{\phi K}$ can be positive only for an $RL$ 
insertion.  A pure $RL$ scenario also predicts a vanishing direct 
CP asymmetry in 
$B\rightarrow X_s \gamma$.  However, if the $RL$ insertion is
accompanied by an $RR$ insertion, 
then the resulting direct CP asymmetry in 
$B\rightarrow X_s \gamma$ can be large \cite{kane}  and the
$RR$ contribution to the $B_s - \overline{B_s}$
mixing can induce significant shifts in $\Delta M_s$ and 
the phase of $B_s - \overline{B}_s$ mixing.  
Finally, we also point out that the $| ( \delta_{LR,RL}^d )_{23} | 
\lesssim 10^{-2}$ can be naturally obtained in SUSY flavor models with 
double mass insertion at large $\tan\beta$, and in 
string-motivated models \cite{kkkpww}. 

\begin{acknowledgments}
We thank S. Baek, H. Davoudiasl, B. Nelson, K. Tobe and J. Wells for 
helpful comments, and A. Kagan for pointing out the sign error in the
corrleation between $S_{\phi K}$ and $A_{\rm CP}^{b\rightarrow s\gamma}$.  
PK and JP are grateful to MCTP at Univ. of Michigan for the hospitality. 
This work is supported in part by 
KRF grant KRF-2002-070-C00022 and 
KOSEF through CHEP at Kyungpook National University (PK and JP), 
by National Science Foundation under grant PHY00--98791 (CK), 
by Department of Energy (GK and HW), 
by a DOE grant No. DE-FG02-95ER40896 and in part by the Wisconsin Alumni
Research Foundation (LW). 
\end{acknowledgments}

\end{document}